\documentclass[prl,twocolumn,superscriptaddress]{revtex4}
\usepackage{graphicx}
\usepackage{amsmath}
\usepackage{dcolumn}
\usepackage{color}
\usepackage{epstopdf}

\newcommand{\bom}{\mbox{\boldmath $\bf\omega$}}

\def\V{\textbf{V}}
\def\v{\textbf{v}}
\def\b{\textbf{b}}
\def\A{\textbf{A}}

\def\j{\textbf{j}}
\def\B{\textbf{B}}

\def\bh{\textbf{h}}

\def\e{\textbf{e}}
\def\pa{\partial}
\def\S{\textbf{S}}
\def\R{\textbf{R}}
\def\r{\textbf{r}}

\begin{document}

\title{Minimum energy states of the plasma pinch in standard and Hall magnetohydrodynamics}

\author{I. V. Khalzov}
\author{F. Ebrahimi}
\author{D. D. Schnack}
\author{V. V. Mirnov}

\affiliation{Center for Magnetic Self-Organization, University of Wisconsin, 1150 University Avenue, Madison, Wisconsin, 53706 USA}

\date{\today}

\begin{abstract}
Axisymmetric relaxed states of a cylindrical plasma column are found analytically in both standard and Hall magnetohydrodynamics (MHD) by complete minimization of energy with constraints imposed by invariants inherent in corresponding models. It is shown that the relaxed state in Hall MHD is the force-free magnetic field with uniform axial flow and/or rigid azimuthal rotation. The relaxed states in standard MHD are more complex due to the coupling between velocity and magnetic field. Application of these states for reversed-field pinches (RFP) is discussed.           
\end{abstract}

\maketitle

Minimum energy states appear in many magnetized plasmas as a result of relaxation --  the spontaneous  tendency to evolve toward preferred configurations with ordered structure. Theoretical prediction of such states is a long-standing problem for both laboratory and astrophysical applications. The concept of relaxation was proposed by Taylor \cite{Taylor}, who conjectured that during turbulent dynamics a slightly resistive magnetohydrodynamic (MHD) system tends to minimize its magnetic energy while conserving the total magnetic helicity. The underlying basis of this approach is the principle of selective decay of invariants \cite{Taylor,Taylor2,Dalton}, i.e.,  one or more ideal invariants of the system (conserved in the absence of dissipation) are less susceptible to dissipation than energy and thus can be considered as constants during the relaxation process. Mathematically  the relaxation theory is formulated as a variational procedure for obtaining a relaxed state by minimizing the energy subject to constraints. 

The Taylor theory is successful in explaining magnetic structures in laboratory plasmas such as the reversed-field pinch (RFP),  multipinch and spheromak \cite{Taylor2,Dalton} but  it does not include flows that are ubiquitous in experimentally observed relaxed states. The origin of these flows is not well understood; laboratory plasmas rotate in the toroidal and poloidal directions even in  the absence of externally applied torques (intrinsic rotation).  Further, the experimental parameters are such that  the single-fluid MHD model may not be strictly valid, and the inclusion of the effects of separate ion and electron fluids in the model may be required.

The present work is motivated by the recent progress in plasma velocity measurements in the Madison Symmetric Torus  RFP experiment, which show an abrupt change of the global flows during the relaxation events. Detailed temporal and spatial measurements of flow  dynamics indicate significant radial angular momentum transport and  flattening of the radial flow profiles \cite{Kuritsyn}.

The goal of the present paper is to determine the  minimum energy (relaxed) states for a cylindrical RFP, to analyze the possibility of plasma flows in such states in  both standard (single-fluid) and Hall MHD (a two-fluid model with massless electrons and uniform density), and to elucidate their global properties. Since the RFP has ``bad''  magnetic curvature everywhere, the geometry of periodic cylinder is a good approximation for RFP theory and simulations \cite{Taylor2,Dalton}. We employ a variational procedure that includes all ideal invariants inherent in corresponding models. While the experiments are open systems that interact with the external environment through applied voltages, here we consider only closed systems.  This is consistent with Taylor's approach \cite{Taylor, Taylor2}, which has been reliable for predicting the general properties of relaxed magnetic fields without flow.  The fields and flows predicted by the present theory  may be relevant to the flows that are observed after the crash phase of sawtooth cycle in the RFP. Of course, we cannot comment on the origin of these flows, only on their relaxed properties. For simplicity, only  axisymmetric states are considered. 

In the framework of incompressible MHD similar studies are reported in Refs. \cite{Chiyoda, comment, Yosh}. In Ref. \cite{Chiyoda} (results are corrected in Ref. \cite{comment}) the cross helicity invariant is included  in the analysis and a relaxed state with velocity parallel to force-free magnetic field is obtained. We emphasize here that the cross helicity in incompressible MHD  is an ideal invariant of the system and is also a rugged invariant in the presence of dissipation, which is confirmed by numerical simulations \cite{Horiuchi}.  In Ref. \cite{Yosh} in addition to the cross helicity the total angular momentum is taken  into account  as a conserved quantity in toroidal geometry. This yields two separate types of relaxed MHD states with force-free magnetic field and either parallel flow or rigid rotation. The novel result of our paper is that the inclusion of the cross helicity along with momenta constraints in cylindrical geometry generally leads to  MHD relaxed state with non force-free magnetic field and mixed types of flows.           

The relaxation problem in the framework of incompressible Hall MHD is considered in Refs. \cite{Turner, Mah,Stein3,Stein4, Yosh2}. In Refs. \cite{Turner, Mah} the double-Beltrami equation for the relaxed states in Hall MHD is  revealed and several solutions are presented for cylindrical geometry.  These solutions do not correspond to true minimum energy states for fixed electron and ion helicities, $H_e$ and $H_i$, since the minimization procedure is not completed. This is because the unknown Lagrange multipliers used in the variational principle are not specified in terms of the initial values of the invariants.  A more complete analysis is reported in Ref. \cite{Stein3} where the energy of relaxed states is found in toroidal systems as a function of $H_e$ and $H_i$.  Although the general solution of the double-Beltrami equation has two eigenfunctions, Ref. \cite{Stein3} uses only one of them.  This precludes the possibility of two different spatial scales in the relaxed state (as in Refs. \cite{Mah, Stein4, Yosh2}).  In Ref. \cite{Stein4} relaxed states are obtained as a linear combination of two orthogonal Beltrami eigenfunctions with eigenvalues $\Lambda_1$ and $\Lambda_2$, respectively, and the energy is expressed as a function of $H_i$, $H_e$, $\Lambda_1$ and $\Lambda_2$. The next step is to find a pair of eigenvalues $(\Lambda_1,~\Lambda_2)$ that minimizes the energy; however, this step is missing in Ref. \cite{Stein4}. 

The full energy minimization of the Hall MHD system with fixed electron and ion helicities is completed  in Ref. \cite{Yosh2}, with the result that the relaxed state in Hall MHD is always a force-free magnetic field  with no plasma flows. The authors of Ref. \cite{Yosh2}  question the conservation of the ion helicity $H_i$, arguing that it is not a rugged invariant during relaxation and, therefore, it should not be included into energy minimization procedure. The fact that it is not conserved in Hall MHD relaxation is confirmed by numerical simulations \cite{Yosh2, Numata1,Numata2}. We adopt these conclusions by  excluding the ion helicity from the invariants 
of the Hall MHD relaxation. Instead, we include other velocity related invariants that  follow from the geometrical symmetry of the system: total axial and angular momenta. This allows us to obtain a relaxed Hall state with plasma flows.



We consider the problem of finding the minimum energy (relaxed) states of cylindrical plasma pinch of length $L$.
We assume that plasma is surrounded by a perfectly conducting shell (flux conserver) of radius $a$, and plasma density $\rho$ is uniform in space and constant in time. Under these assumptions the plasma is described by equations of ideal incompressible Hall MHD, which in non-dimensional form are \cite{Turner}
\begin{eqnarray}
\label{HMHD1}\frac{\pa\v}{\pa\tau}&+& (\v\cdot\nabla)\v+\nabla p = (\nabla\times\b)\times\b,~~\nabla\cdot\v=0,\\
\label{HMHD2}\frac{\pa\b}{\pa\tau}&=&\nabla\times(\v\times\b-\delta_i(\nabla\times\b)\times\b),~~\nabla\cdot\b=0.
\end{eqnarray} 
Here, all physical quantities are normalized:  
$$
\R=a\r,~~~t=\frac{a}{V_A}\tau,~~~\V=V_A\v,~~~\B=B_0\b,~~~P=\rho V_A^2p,
$$
where the Alfv\'{e}n velocity $V_A$ and the characteristic magnetic field $B_0$ are defined as
\begin{equation}\label{Phi_0}
V_A=\frac{B_0}{\sqrt{4\pi\rho}},~~~B_0=\frac{\Phi_0}{\pi a^2},
\end{equation}
with  $\Phi_0$ being a total axial magnetic flux. System (\ref{HMHD1})-(\ref{HMHD2}) also contains the non-dimensional ion skin depth (or Hall parameter) $\delta_i$, which is  defined as   
$$
\delta_i=\frac{d_i}{a}=\frac{c}{a}\sqrt{\frac{m_i^2}{4\pi\rho e^2Z^2}},
$$
where $m_i$ and $eZ$ are  ion mass and charge, $c$ is speed of light. In the case $\delta_i=0$, single-fluid MHD is recovered. 

We adopt cylindrical coordinate system $\{r,\varphi,z\}$. In order to solve Eqs. (\ref{HMHD1}), (\ref{HMHD2}) uniquely, we have  to specify boundary conditions for velocity and magnetic field. At a perfectly conducting, impermeable surface the normal components of the velocity and  the magnetic field vanish,
\begin{equation}
\label{BC_n} v_n=0,~~~b_n=0, 
\end{equation}
and for closed systems the tangential component of electric field is zero, which is equivalent to 
\begin{equation}
\label{BC_Et}  \delta_i(\j_n\times\b_t) =0 ~~~\textrm{or}~~~\delta_ij_n=0, 
\end{equation}
where $\j=\nabla\times\b$ is normalized current density. Note that in single-fluid MHD ($\delta_i=0$) this condition is satisfied automatically. 


Following the fundamental idea of the Taylor theory, we introduce a number of  global invariants and examine them  for cylindrical pinch geometry assuming boundary conditions (\ref{BC_n}), (\ref{BC_Et}). The total energy of system (\ref{HMHD1}), (\ref{HMHD2}),
\begin{equation}
\label{E} E=\frac{1}{2}\int(\v^2+\b^2)d^3\r,
\end{equation}
is an ideal invariant.  In a dissipative isolated plasma, total energy  can only decrease in time. This validates the procedure of energy minimization in relaxation theory.  


Magnetic helicity is equivalent to electron helicity $H_e$ for massless electrons:  
\begin{equation}
\label{K} h_1\equiv H_e =\int \A\cdot\nabla\times\A d^3\r,
\end{equation}
where $\A$ is vector potential, $\b=\nabla\times\A$. Magnetic helicity is an ideal invariant; it changes only in presence of resistivity. For our study it is important that magnetic helicity is more robust than the energy  \cite{Taylor,Taylor2,Dalton}, i.e., its value is approximately constant in time and equal to its initial value, $h_1=K$.  


As it has been mentioned above the ion helicity
\begin{equation}
\label{Ki}
H_i=\int(\A+\delta_i\v)\cdot\nabla\times(\A+\delta_i\v) d^3\r 
\end{equation}
is not a good invariant in Hall MHD. To illustrate this more precisely we introduce generalized cross helicity as
\begin{equation}
\label{M} h_2=\int\v\cdot(\b+\frac{\delta_i}{2}\nabla\times\v)d^3\r,
\end{equation}
which is related to ion helicity by the equation:
\begin{equation}
      H_i=H_e+2\delta_i h_2 +\delta_i\int_S(\v\times\A)\cdot d\S, \nonumber
\end{equation}
where the last integral is taken over the cylindrical  surface of the plasma.  
Though the time derivative of the integrand in Eq. (\ref{M}) is reduced to full divergence, the generalized cross helicity (and, therefore, the ion helicity) is not an ideal invariant in Hall MHD, since  
\begin{equation}
\label{M_t} \frac{\pa h_2}{\pa\tau}=\frac{\delta_i}{2} \int_S  \bigg(\frac{\v^2}{2}-p\bigg)\bom\cdot d\S\ne0,
\end{equation}
where $\bom=\nabla\times\v$ is the fluid vorticity. In order for $h_2$ to be conserved, an extra boundary condition must be  imposed, which is $\omega_n=0$  \cite{Turner}. However, this boundary condition overspecifies the solution to Eqs. (\ref {HMHD1}) and (\ref{HMHD2}). Thus, the ruggedness of the ion helicity  should not be assumed in Hall relaxation.

If $\delta_i=0$, Eq. (\ref{M}) defines the MHD cross helicity:  
\begin{equation}
\label{H2} h_2=\int\v\cdot\b d^3\r.
\end{equation}
This quantity is an ideal invariant in incompressible MHD. Moreover, cross helicity (\ref{H2}) is also a rugged invariant in MHD relaxation, which is confirmed by numerical simulations \cite{Horiuchi}. In our MHD analysis  we assume that the  cross helicity is constant and equal to its initial value, $h_2=M$.       


The perfectly conducting cylindrical shell serves as a conserver of axial magnetic flux: 
\begin{equation}
\label{H3} h_3=\int\limits_0^{2\pi}\int\limits_0^1 b_z rdrd\varphi.
\end{equation}
This is the only true invariant in the system, which is conserved in both ideal and dissipative plasmas. Based on normalization (\ref{Phi_0}),  $h_3=\pi$.


The symmetry of the pinch configuration yields two more ideal invariants, the axial and angular momenta:
\begin{eqnarray}
\label{H4} h_4&=&\int\v\cdot\e_zd^3\r,\\
\label{H5}h_5&=&\int r\v\cdot\e_\varphi d^3\r.
\end{eqnarray}
The initial values of these invariants can always be attributed to some uniform flow with velocity $u$ in $z$-direction  and a rigid rotation with angular velocity $\Omega$  in $\varphi$-direction, i.e., $h_4=\pi lu$ and $h_5=\pi l\Omega/2$, where $l=L/a$ is the normalized length of the cylinder.   


We look for a minimum of energy (\ref{E}) subject to  constraints given by  global invariants (\ref{K}), (\ref{H2})-(\ref{H5}). The cross helicity (\ref{H2}) is used as a constraint only in standard MHD but not in Hall MHD. The Lagrange multipliers method gives us a conditional extremum of $E$ (it is minimum since $E$ is positive definite, and, therefore, the resulting equilibrium is ideally stable): 
\begin{eqnarray}
\label{main1}&&\v_0= - \mu_2\b_0-\mu_4\e_z-\mu_5r\e_\varphi ,\\
\label{main2}&&(1-\mu_2^2)\nabla\times\b_0+ 2\mu_1\b_0=2\mu_2\mu_5\e_z,\\
\label{main3}&&h_1=K, h_2=M, h_3=\pi, h_4=\pi lu, h_5=\frac{\pi l\Omega}{2}.
\end{eqnarray}
Note that the amplitude of the magnetic field $\b_0$ and the Lagrange multipliers $\mu_1$, $\mu_2$, $\mu_4$, $\mu_5$ are not arbitrary, they are determined by the constraints (\ref{main3}). Therefore, the relaxed state depends only on the initial values of the invariants. 

Eqs. (\ref{main1})-(\ref{main3}) describe single-fluid MHD relaxed state. The Hall MHD relaxed state is obtained  from these equations by setting $\mu_2=0$ and ignoring cross helicity constraint $h_2=M$. The axisymmetric Hall state is 
\begin{eqnarray}
\label{hall0}\A_0&=&\frac{1}{2J_1(\lambda)}\,\bigg(\bh(\lambda r) - J_0(\lambda)\e_z \bigg),\\
\label{hall1}\b_0&=&\frac{\lambda}{2J_1(\lambda)}\,\bh(\lambda r),\\
\label{hall2}\v_0&=&\Omega r\e_\varphi + u \e_z,
\end{eqnarray} 
where $\lambda=-2\mu_1$,
\begin{equation}
\label{bh}
\bh(\lambda r)=J_1(\lambda r)\e_\varphi + J_0(\lambda r)\e_z,~\nabla\times\bh=\lambda\bh,
\end{equation}
and $J_0$ and $J_1$ are Bessel functions  of the first kind.  For unique definition of vector potential $\A_0$ we take a gauge invariance condition \cite{Reiman2}   
\begin{equation}\label{gauge}
\int\limits_0^l A_z\bigg |_{r=1}dz=0.
\end{equation}
Solution given by Eqs. (\ref{hall0})-(\ref{hall2}) is in agreement with Ref. \cite{Yosh2}; it corresponds to a force-free Taylor state \cite{Taylor,Taylor2,Dalton} for magnetic field with addition of  rigid rotation in azimuthal direction or/and uniform flow in axial direction.  Parameter  $\lambda$ is determined from the magnetic helicity:
\begin{equation}\label{Tay_h1}
\tilde{K}\equiv\frac{K}{\pi l}=\frac{\lambda}{2}\bigg(1-\frac{J_0(\lambda)J_2(\lambda)}{J_1^2(\lambda)} \bigg).
\end{equation} 
Here and below we use tilde to denote the same quantity divided by $\pi l$, i.e., the normalized density of the quantity.
For values of magnetic helicity larger than $\tilde{K}>4.1$  the minimum energy state becomes helically symmetric \cite{Reiman}.       

Next we consider the single-fluid MHD relaxed state. In this case a solution to system (\ref{main1})-(\ref{main3}) is 
\begin{eqnarray}
\label{mhd0}\A_0&=&\frac{C}{\lambda}\,\bigg(\bh(\lambda r) - J_1(\lambda)r\e_\varphi - J_0(\lambda)\e_z \bigg) + \frac{r}{2} \e_\varphi,\\
\label{mhd1}\b_0&=&C\,\bigg(\bh(\lambda r) - \frac{2J_1(\lambda)}{\lambda}\e_z \bigg) + \e_z,\\
\label{mhd2}\v_0&=&-\mu_2C\,\bigg(\bh(\lambda r) - \frac{4J_2(\lambda)}{\lambda}\,r \e_\varphi  - \frac{2J_1(\lambda)}{\lambda}\e_z \bigg) \\
 &+& \Omega r\e_\varphi + u \e_z, \nonumber
\end{eqnarray} 
where $\bh$ is given by Eq. (\ref{bh}), $\lambda$ and the amplitude $C$ are
\begin{equation}\label{ampl}
\lambda=\frac{2\mu_1}{\mu_2^2-1},~~~C=\frac{[\lambda(1-\mu_2^2) - 2\Omega\mu_2]}{2[J_1(\lambda) +\mu_2^2J_3(\lambda)]}. 
\end{equation}
Substituting these expressions into Eqs. (\ref{K}) and (\ref{H2}) we obtain the magnetic helicity and the cross heicity:
\begin{eqnarray}
\label{MH}\tilde{K}&=&\frac{2CJ_2(\lambda)}{\lambda} \\
&+&\frac{2C^2}{\lambda}\bigg(J_1^2(\lambda) - 2J_0(\lambda)J_2(\lambda) - J_2^2(\lambda) \bigg),\nonumber\\
\label{CH}\tilde{M}&=&u + \frac{2\Omega C J_2(\lambda)}{\lambda}\\
 &-& \mu_2C^2\bigg(2J_1^2(\lambda) - 3J_0(\lambda)J_2(\lambda) - J_2^2(\lambda) - \frac{8J_2^2(\lambda)}{\lambda^2} \bigg).\nonumber
\end{eqnarray} 
Eqs. (\ref{MH}) and (\ref{CH}) are used to find  $\mu_2$ and $\lambda$ through the initial values of helicities, $\tilde{K}$ and $\tilde{M}$. In the limit $\mu_2\to0$, Eqs. (\ref{mhd0})-(\ref{mhd2}) and (\ref{MH}) become Eqs. (\ref{hall0})-(\ref{hall2}) and (\ref{Tay_h1}), respectively. We also note here that the uniform axial flow $u$ can be removed from all equations. This corresponds to transformation to the reference frame moving with velocity $u$ in $z$ direction: 
$$
\v^*_0=\v_0-u\e_z,~\tilde{M}^*=\tilde{M}-u.
$$   
Without loss of generality we assume that $u=0$. 

\begin{figure}[tbp]
\begin{center}
  \includegraphics[scale=0.4]{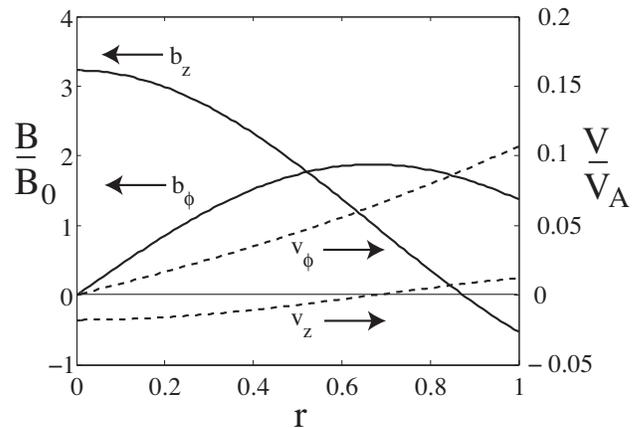}\\
  \caption{MHD relaxed state with magnetic helicity $\tilde{K}=2$,  cross helicity $\tilde{M}=0.1$ and angular momentum $\Omega=0.1$. Solid lines are components of magnetic field (left vertical axis), dashed lines are components of velocity (right vertical axis).}\label{fig1}
\end{center}
\end{figure}

\begin{figure}[tbp]
\begin{center}
  \includegraphics[scale=0.4]{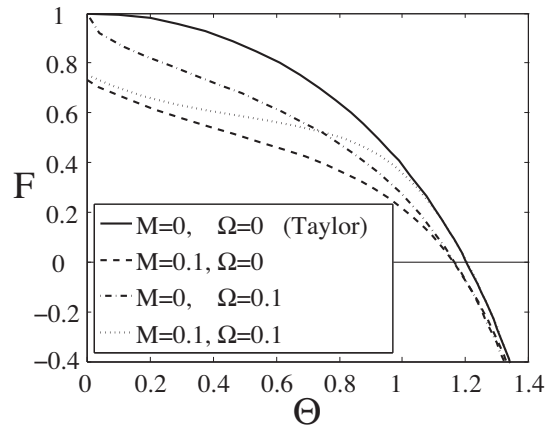}\\
  \caption{Reversal parameter $F$ vs. pinch $\Theta$ for the relaxed MHD states  with  different values of cross helicity $\tilde{M}$ and angular momentum $\Omega$.}\label{fig2}
\end{center}
\end{figure}

Our main results are summarized in Figs. \ref{fig1}  and \ref{fig2}. A sample of MHD relaxed state (\ref{mhd1}), (\ref{mhd2}) with non-zero plasma flow is presented in Fig. \ref{fig1}. Fig. \ref{fig2} shows the $F-\Theta$ diagram of the relaxed MHD states with different values of cross helicity $\tilde{M}$ and  angular momentum $\Omega$. The reversal parameter $F$ and pinch  $\Theta$ are  defined here as 
\begin{eqnarray}
\label{F} F&=&\frac{B_z|_{R=a}}{B_0}=b_z|_{r=1},\\
\label{Theta} \Theta&=&\frac{B_{\varphi}|_{R=a}}{B_0}=b_\varphi|_{r=1}.
\end{eqnarray} 
As follows from Fig. \ref{fig2}, the presence of the initial flow (non zero cross helicity or total angular  momentum) in cylindrical plasma pinch affects the relaxed magnetic field significantly. This is due to the coupling  of the velocity and the magnetic field that occurs in Eq. (\ref{main2}) through the term on the right hand side. Such coupling does not take place in the systems without axial symmetry where  the angular momentum (\ref{H5}) is not conserved, e.g., in a periodic rectangular box. In this case a relaxed  magnetic field corresponds to a  force-free Taylor state  and it has the same structure in both standard and Hall MHD, independent of the initial flows and determined by the value of magnetic helicity only \cite{Numata1, Numata2}.

In summary, we have found the axisymmetric minimum energy (relaxed) states of a cylindrical plasma pinch within the framework of both standard and Hall MHD. Like all variational theories of plasma relaxation, the present calculation is silent as to the details of the dynamics that are responsible for the relaxation process.  The only requirement is that they preserve the robust invariants assumed during the variational procedure. Relaxation with Hall MHD does not preserve the generalized cross helicity, leads to states with uniform axial flow and constant angular velocity, and reproduces the well-known Bessel-function model for the magnetic field profiles. Relaxation with single-fluid MHD robustly preserves the cross helicity and leads to flow profiles with more radial structure, but it also modifies the form of the relaxed magnetic field. In both cases the solutions predict redistribution (transport) of the  momentum caused by plasma relaxation in a closed system. This conclusion is qualitatively consistent with flow dynamics observed in RFP experiment.  At this time there is insufficient experimental information to distinguish between  MHD and Hall MHD relaxation.   Further insight in this regard requires more detailed experimental measurements, and full nonlinear Hall MHD computations that take into account the driven (open) character of RFP system.

The authors  wish to thank C. Hegna and C. Sovinec for useful discussions. This work is supported by  the National Science Foundation and by the U.S. Department of Energy under Grant DE-FG02ER-54868.

\end{document}